\documentstyle[prl,aps,multicol,epsf,rotate]{revtex}
\begin{document}
\draft
\preprint{XXXX}
\title{Linking numbers for self-avoiding loops and percolation:\\ 
application to the spin quantum Hall transition}
\author{John Cardy}
\address{University of Oxford, Department of Physics -- Theoretical
         Physics, 1 Keble Road, Oxford OX1 3NP, U.K. \\
         and All Souls College, Oxford.}
%
%\date{\today}
%
\maketitle
\begin{abstract}
Non-local twist operators are introduced
for the ${\rm O}(n)$ and $Q$-state Potts models
in two dimensions which, in the limits $n\to0$ (resp.~$Q\to1$) 
count the numbers of self-avoiding loops (resp.~percolation clusters)
surrounding a given point. Their scaling dimensions are
computed exactly. This yields many results, for example
the distribution of the number of percolation clusters
which must be crossed to connect a given point to an infinitely distant
boundary. Its mean behaves as
$(1/3\sqrt3\pi)|\ln(p_c-p)|$ as $p\to p_c-$.
These twist operators correspond to $(r,s)=(1,2)$ in the
Kac classification of conformal field theory, so that their higher-order
correlation functions, which describe linking numbers around multiple
points, may be computed exactly. 
As an application we compute the exact value $\sqrt3/2$ for the
dimensionless conductivity at the spin Hall transition, as well
as the shape dependence of the mean conductance in an arbitrary simply
connected geometry with two extended edge contacts.
\end{abstract}
\begin{multicols}{2}
The conformal field theory/Coulomb gas
approach to two-dimensional percolation and
self-avoiding walk problems has been extraordinarily fruitful
\cite{Nien,DotFat,diFran}. In
addition to values for many of the critical exponents, other
universal scaling functions such as percolation crossing probabilities
have been obtained exactly \cite{JCcrossing,Wat}. 
In this Letter a set of
correlation functions of non-local operators is introduced,
which describe topological properties of percolation clusters
and self-avoiding walks, and count the number of
clusters or loops which must be crossed in order to connect two or more
points. 

It turns out that these exponents may easily be computed using standard
Coulomb gas methods, for general $n$ or $Q$ in the ${\rm O}(n)$ or $Q$-state
Potts model respectively. In the limits $n\to0$ or $Q\to1$,
corresponding to self-avoiding walks or to percolation respectively,
the scaling dimensions of these operators vanish, so that some of their
correlations are trivial. However, it is their
\em derivatives \em with respect to $n$ or $Q$ which give physical information,
thereby giving rise to a variety of logarithmic behavior. The frequent
occurrence of logarithmic correlations in such conformal field
theories (CFTs) with vanishing central charge has recently been pointed
out in several contexts \cite{logs,GurLud}. 

In certain cases these topological operators may be recognized as
\em twist \em operators which have already been identified in $c=1$
theories \cite{Saltwist} and the 3-state Potts model \cite{JCtwist}. 
{}From the CFT point of view,
they turn out to correspond to degenerate Virasoro representations 
labeled by $(r,s)=(1,2)$ in the Kac classification. These bulk operators
have not previously been identified for general $n$ and $Q$. The fact
that they are degenerate means that their higher-point correlations may
be computed exactly.

In addition to bulk operators of the above type it is also possible to
define operators which count the number of loops or clusters
surrounding a point
near the boundary of a system. They
may be used to compute the universal mean 
conductance at the spin quantum Hall transition, which has recently
been shown to map onto the percolation problem \cite{Gruz}.

\noindent\em ${\rm O}(n)$ model\em. 
Let us recall the elements of the Coulomb gas approach
\cite{Nien}.
$n$-component spins ${\bf s}(r)$ with
${\bf s}^2(r)=1$ are placed at the sites $r$ of a lattice, with a
nearest neighbor interaction. The partition
function ${\rm Tr}\,\prod_{rr'}\big(1+y{\bf s}(r)\cdot{\bf s}(r')\big)$,
when expanded in powers of $y$, gives a sum over self-avoiding loops with
a factor $y$ for each bond and $n$ for each loop. Each loop may be replaced
by a sum over its orientations, with a weight $e^{\pm i\pi\chi}$ for each,
with $\chi$ chosen so that the sum gives $n=2\cos\pi\chi$.
This gas of oriented loops is then mapped onto a height model with
variables $\phi(R)\in\pi{\rm {\bf Z}}$ on the dual lattice, such that 
on the dual bond $RR'$
$\phi(R)-\phi(R')=0,\pm1$ according to whether the bond it crosses is
empty or is occupied by a loop segment of one orientation or the other.
At the critical fugacity $y_c$
this is supposed to renormalize onto a free field with reduced free
energy functional $(g/(4\pi))\int(\partial\phi)^2d^2\!r$, with respect
to which the long-distance behavior of all correlations may be computed
as long as phase factors associated with non-contractible loops are
correctly accounted for. $g$ may be fixed by a variety of methods: for
the dilute regime of interest here, $g=1+\chi$. 
On an infinitely long cylinder of perimeter $L$, 
in order to correctly count loops
which wrap around the cylinder with weight $n$ 
it is necessary to insert factors 
$e^{\pm i\chi/\phi(\pm\infty)}$ at either end: these modify the
free energy per unit length to $-(\pi/6L)(1-(6/g)\chi^2)$, from which
the value of the central charge $c=1-6(g-1)^2/g$ follows. For
self-avoiding walks, $n=0$, $\chi=\frac12$, and $g=\frac32$, so that
$c=0$. 

Now suppose that loops which wrap around the cylinder are counted with a
different weight $n'=2\cos\pi\chi'$. (A similar construction was used in
Ref.~\cite{DSwinding} to count the winding angle of open walks).
The cylinder free energy per unit
length will be modified by a term $(\pi/g)({\chi'}^2-\chi^2)$. In the
plane, this may be interpreted as the insertion of a non-local operator 
$\sigma_{n'}$ whose scaling dimension is
\begin{equation}
\label{xnn}
x(n;n')=(1/2g(n))\big({\chi'}^2-\chi^2\big).
\end{equation}
so that the two-point function
$\langle\sigma_{n'}(r_1)\sigma_{n'}(r_2)\rangle$ decays like
$|r_1-r_2|^{-2x(n;n')}$ at criticality. The interpretation of this is as
a \em defect line \em joining $r_1$ and $r_2$: loops which cross this
defect line an odd number of times, that is, which surround either but
not both of the points\cite{sphere},
now carry a factor $n'$ instead of $n$. A particular example is
$n'=-n$, which is equivalent to making
the identification ${\bf s}(r)\to-{\bf s}(r)$ as the defect line is
crossed. Such \em twist \em operators have been identified in other
models\cite{Saltwist,JCtwist}: 
like disorder operators, they reflect the existence of
non-trivial homotopy.
{}From (\ref{xnn}) it follows that the dimension of
this operator is $x(n;-n)=(3/2g)-1$, 
which is $x_{1,2}$ in the Kac classification
$x_{r,s}=\big((rg-s)^2-(g-1)^2\big)/2g$. However, for small
$n$ and $n'$, 
$x(n,n')\sim(1/6\pi)(n-n')$, so that, to first order,
it does not matter whether $x(n,-n)$ or $x(0;n')$ is used apart from
factors of 2. Note that if $n'=-n$ the fusion rules of CFT
imply the operator product expansion (OPE)
$\sigma\cdot\sigma=1+\epsilon+\cdots$, where $\epsilon$ is the
$(1,3)$ operator which has been identified as the local energy density of the 
${\rm O}(n)$ model \cite{DotFat}. $\epsilon(r)$ counts whether
a given bond $r$ is occupied or not -- consistent with 
the interpretation that,
as $a\to0$, the $O(n)$ term in
$\sigma(r+a)\sigma(r-a)$ counts whether the single loop is
trapped between the points $r\pm a$.

\begin{figure}[tbp]
\centerline{\epsfxsize=3in \epsfbox{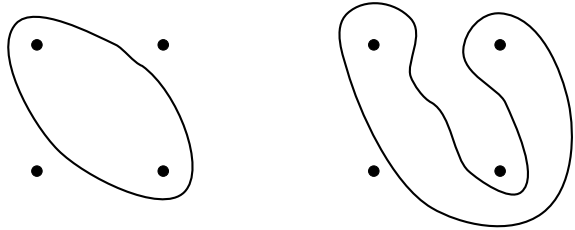}}
\caption{Examples of loops which wind non-trivially around 4 points,
counted by the connected 4-point function
$\langle\sigma\sigma\sigma\sigma\rangle_c$.}
\label{fig1}
\end{figure}

Here are a few examples of the application of this formula. 
Away from criticality it implies that
$\langle\sigma_{n'}(R)\rangle\sim C(n;n')\xi^{-x(n,n')}$ where the
correlation length $\xi$ behaves as $(y_c-y)^{-\nu}$, and $C(0;0)=1$.
When $n=0$, the coefficient of
${n'}^M$ counts configurations of $M$ nested loops surrounding the point
$R$ of the dual lattice.
Denoting the number of such loops of \em total \em length $N$ by
$b_N^{(M)}$, the most singular term in the generating function
$\sum_Nb_N^{(M)}y^N\sim(1/M!)((1/8\pi)|\ln(y_c-y)|)^M$ as
$y\to y_c-$, using $\nu=\frac34$.
This yields the asymptotic behavior
\begin{displaymath}
b_N^{(M)}\sim (\sigma_0/(M-1)!)(1/8\pi)^M(1/N)(\ln N)^{M-1}\mu^N, 
\end{displaymath}
where $\mu=y_c^{-1}$
is the usual lattice-dependent connective constant,
and $\sigma_0$ is a positive integer taking account of the fact that, on
a loose-packed lattice, there are $\sigma_0$ equivalent
singularities on the circle $|y|=y_c$. 
For $M=1$, one may sum over the position of the marked point $R$, instead
of summing over the position of the loop, so that
$b_N^{(1)}=A_Np_N$, where $p_N$ is the number of single loops
per lattice site and $A_N$ is their average area. The amplitude $1/8\pi$
in this case agrees with an earlier result found by a different (more
involved) method \cite{JCarea}.
It is interesting also to calculate higher point functions. 
For example the $O(n)$ term in the \em connected \em 4-point function 
$\langle\sigma\sigma\sigma\sigma\rangle_c$
counts the number of single loops which have non-trivial winding around
all four points. Examples are shown in Fig.~\ref{fig1}. Since $\sigma$
is a (1,2) operator this 4-point function may be computed exactly at
criticality, in terms of hypergeometric functions. The details will be
given elsewhere \cite{JCinprep}.

\noindent\em Percolation\em. The Coulomb gas formulation of the
$Q$-state Potts model is similar to the above, except that it is
valid only at the critical point \cite{Nien}. 
The model is first mapped onto the
random cluster model, in which every cluster configuration
is weighted by a factor of $p/(1-p)$ for each bond and $Q$ for each
cluster. Each cluster may be identified by its outer and inner hulls,
which form a dense set of closed loops. At criticality, each hull then
carries a weight $\sqrt{Q}$. Once again this may be mapped onto a gas
of oriented loops with phase factors $e^{\pm\pi\chi}$, where 
$\sqrt{Q}=2\cos\pi\chi$, and then to a free field theory, where,
however, this time $g=1-\chi$. The central charge then
vanishes for $g=\frac23$, corresponding to
$\chi=\frac13$ or $Q=1$, the percolation limit. Once again one may define
a non-local operator which counts the hulls which surround a marked
point with a different weight $\sqrt{Q'}=2\cos\pi\chi'$, and standard
Coulomb gas methods then lead to a result identical in form to 
(\ref{xnn}) for its dimension $x(Q;Q')$. 
For this to be given by $x_{(1,2)}=(3g/2)-1$, 
$Q'=(2-Q)^2$. For $Q=3$, for example, the clusters which wrap
around the marked point are counted with weight $Q'=1$. This is
consistent \cite{nonleading} with the identification of the ${\rm\bf Z}_2$
twist operator for the 3-state
model made in Ref.~\cite{JCtwist}.
For the Ising model ($Q=2$), it is the same as the disorder operator. 
Note that as $Q\to1$, the $O(Q-1)$ term in $\sigma(r+a)\sigma(r-a)$
counts the \em number \em of clusters separating $r\pm a$, as compared
with the ${\rm O}(n)$ model where it simply counted whether or not the
single loop passed between them. For this reason the lattice
interpretation of the (1,3) operator in the OPE $\sigma\cdot\sigma$
is problematic in this case\cite{AAKK}.

For $Q=1+\delta$ and $Q'=1+\delta'$ ($|\delta|,|\delta'|\ll1$) note that
$x(Q;Q')\sim(1/4\sqrt3\pi)(\delta-\delta')$ (with
$\delta'\sim-2\delta$ corresponding to $x_{1,2}$),
while for small $Q'$ 
$x(1;Q')=\frac5{48}-\frac3{8\pi}\sqrt{Q'}+O(Q')$. As before, one
application of these results is to count the number of clusters which
surround a given point, that is, which have to be crossed to escape to
infinity. If $P(M)$ is the probability of there being exactly $M$ such
clusters,
\begin{displaymath}
\sum_MP(M){Q'}^M\sim C(Q')(p_c-p)^{\nu x(1;Q')}
\end{displaymath}
where now $\nu=\frac43$.
For $Q'=0$ this gives $P(0)$, which is the probability that the dual
site $R$ is connected to the boundary by a set of dual bonds:
this gives the usual exponent $\beta=\frac5{36}$ of percolation.
Expanding in powers of $\sqrt{Q'}$ now gives
\begin{displaymath}
P(M)\sim P(0)\big(|\ln(p_c-p)|/2\pi\big)^{2M}/(2M)!
\end{displaymath}
Note that although the amplitude in $P(0)$ is not universal, the
ratios $P(M)/P(0)$ are.

However, this is valid only for $M\ll|\ln(p_c-p)|$. To find the
behavior near the average value, expand around $Q'=1$. 
The mean value $\overline M\sim (1/3\sqrt3\pi)|\ln(p_c-p)|$, and the
variance $\overline{M^2}-{\overline M}^2\propto|\ln(p_c-p)|$ also.
This implies that, as $p\to p_c-$, the distribution of $M$ becomes
peaked about $\overline M$.
Alternatively, one could work at the percolation threshold in a finite
system of size $L$, in which case $\overline M\sim(1/4\sqrt3\pi)\ln L$.

\noindent\em Spin quantum Hall transition\em. Recently Gruzberg et
al.\cite{Gruz} have shown that certain properties of a model of
non-interacting quasiparticles for the spin Hall transition (a 2D
metal-insulator transition in a disordered system in which time-reversal
symmetry is broken but ${\rm SU}(2)$ spin-rotation symmetry is not 
\cite{Zirn}) may be mapped exactly onto percolation. In particular, the
mean conductance between two extended contacts on the boundary of a finite
system is (apart from a factor of 2 for the spin sum) equal to the
mean number of distinct clusters whose outer hulls connect the two contacts. As 
argued above, such quantities are related to the derivative with
respect to $Q$ at $Q=1$ of correlation functions of a twist operator.
This will however now be a \em boundary \em twist operator. While it is
possible to adapt the above Coulomb gas arguments to account for the
boundary, since in other examples such methods are known to fail for
boundary operators, we shall instead give a more direct argument. 

\begin{figure}[tbp]
\centerline{\epsfxsize=4in \epsfbox{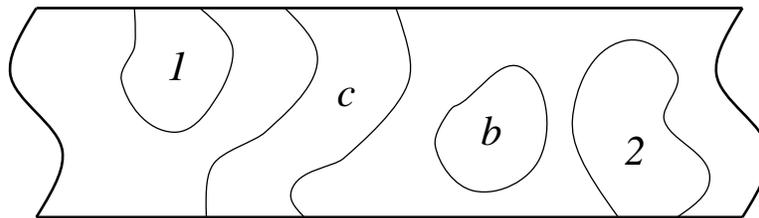}}
\caption{Annular geometry with contacts along either edge. Periodic
boundary conditions are implied in the horizontal direction. Examples
are shown of Potts clusters of types $c$, $1$, $2$ and $b$. The
mean conductance is proportional to the mean number of type $c$.}
\label{fig2}
\end{figure}

First consider the example of an annulus of width $L$ and
circumference $W$. The geometry is shown in Fig.~\ref{fig2}. Apart from
the clusters whose outer hulls 
cross the sample, there are those which touch the
lower edge but not the upper, those which do the opposite, 
and those which touch neither
edge. There may also be one cluster which crosses the sample but which
also wraps around the annulus, so that its outer hulls do not connect
the contacts.
Denote the numbers of such clusters in a given configuration of
the random cluster version of the Potts model by $N_c$, $N_1$, $N_2$, 
$N_b$ and $N_w$ respectively. (Note that $N_c=0$ if $N_w=1$).
Let $Z_{ij}(Q)$ denote the Potts model partition
function with boundary condition of type $i$ on the lower edge and $j$
on the upper edge. The cases of interest are where $i$ or $j$
correspond to either free boundary conditions, denoted by $f$, or to
fixed, in which the Potts spins on the boundary are frozen
into a given state, say $1$. Then
\begin{eqnarray*}
Z_{ff}&=&\langle Q^{N_c+N_w+N_1+N_2+N_b}\rangle\qquad
Z_{11}=\langle Q^{N_b}\rangle\\
Z_{1f}&=&\langle Q^{N_2+N_b}\rangle\qquad\qquad
Z_{f1}=\langle Q^{N_1+N_b}\rangle\\
\end{eqnarray*}
so that 
\begin{equation}
\label{Ncav}
\langle N_c+N_w\rangle=(\partial/\partial Q)|_{Q=1}
(Z_{ff}Z_{11}/Z_{f1}Z_{1f})
\end{equation}
According to the theory of boundary CFT \cite{JCbound}, 
$Z_{ij}\sim \exp\big[\pi((c/24)-\Delta_{ij})(W/L)\big]$ as $W/L\to\infty$, 
where $\Delta_{ij}$ is the lowest scaling dimension out
of all the conformal blocks
which can propagate around the annulus with the given boundary
conditions. When $i=j$ this corresponds to the
identity operator, so that $\Delta_{ii}=0$, but for the mixed case
$(ij)=(f1)$ it corresponds to the (1,2) Kac operator, so that $\Delta_{f1}=
\Delta_{1,2}=
\frac12x_{1,2}(Q)$ in the previous notation. This identification was
previously used \em at \em $Q=1$
in Ref.~\cite{JCcrossing} to compute crossing probabilities, i.e. the
probability that $N_c>0$, in simply connected regions.
Substituting into (\ref{Ncav}) gives
$\langle N_c\rangle\sim 2\pi\Delta_{1,2}'(1)(W/L)$ as $W/L\to\infty$, 
since $N_w\leq1$. From this follows
the universal critical conductivity $\sqrt3/2$. 

\begin{figure}[tbp]
\centerline{\epsfxsize=2in \epsfbox{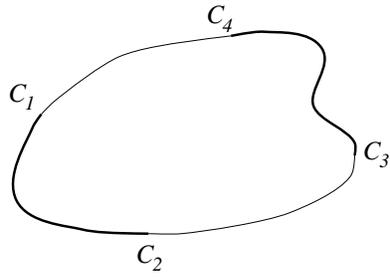}}
\caption{Simply connected region with contacts $C_1C_2$
and $C_3C_4$ along its edge. }
\label{fig3}
\end{figure}

At finite $W/L$ the corrections to the mean conductance
are expected to be of the order of $e^{-\pi\Delta_{2,2}W/L}$
where $\Delta_{2,2}=\frac18$ at $Q=1$, but the full dependence
requires knowledge of the entire operator content of the model for the
different boundary conditions. This seems to be beyond the reach of
current methods.
However, for a \em simply connected \em finite sample the
arguments of Ref.~\cite{JCcrossing} may be adapted. Consider a simply
connected region with contacts $C_1C_2$ and $C_3C_4$ on its boundary, as
shown in Fig.~\ref{fig3}. The remainder of the boundary has hard wall
conditions on the quasiparticle wave functions, corresponding to free boundary
conditions on the Potts spins. The mean number of clusters crossing
between the contacts is still given by (\ref{Ncav}) (with $N_w=0$), 
where the different
boundary conditions are placed on the segments $C_1C_2$ or $C_3C_4$,
with the remaining boundary Potts spins being free. 
This may then be written in terms
of correlation functions of boundary condition 
changing operators\cite{JCbound}
\begin{displaymath}
\langle N_c\rangle={\partial\over\partial Q}\Big|_{Q=1}
\left({\langle
\phi_{f1}(C_1)\phi_{1f}(C_2)\phi_{f1}(C_3)\phi_{1f}(C_4)\rangle\over
\langle\phi_{f1}(C_1)\phi_{1f}(C_2)\rangle
\langle\phi_{f1}(C_3)\phi_{1f}(C_4)\rangle}\right)
\end{displaymath}
These correlation functions are computed by conformally mapping the
interior of the region to the upper half plane. Any conformal rescaling
factors for $Q\not=1$ cancel in the ratio, which then depends only on
the cross-ratio $\eta=(z_1-z_2)(z_3-z_4)/(z_1-z_3)(z_2-z_4)$
of the images $z_i$ of the points $C_i$ under this
mapping. For a rectangle with $|C_1C_2|=W$ and $|C_2C_3|=L$, 
$\eta=(1-k)^2/(1+k)^2$ where $W/L=K(1-k^2)/2K(k^2)$ and $K$ is the
complete elliptic integral of the first kind.  
Since $\phi_{f1}$ is a degenerate (1,2) operator, its four-point
function satisfies a hypergeometric equation. The details of this
calculation will be given elsewhere \cite{JCinprep}. 
The result for the mean conductance is 
\begin{displaymath}
\bar g = 1-{\sqrt3\over2\pi}\left(\ln(1-\eta)+
2\sum_{m=1}^\infty{(\textstyle{1\over3}\displaystyle)_m\over
(\textstyle{2\over3}\displaystyle)_m}{(1-\eta)^m\over m}\right)
\end{displaymath}
For $W/L\gg1$ this reproduces the above result for the conductivity.
In the opposite limit $\bar g\sim Ae^{-(\pi/3)(L/W)}$, in agreement
with Ref.~\cite{Gruz}, but now with a definite prefactor
$A=3\Gamma(\frac23)/2\Gamma(\frac13)^2$.

The author thanks J.~Chalker, A.~Ludwig, V.~Gurarie
and H.~Saleur for useful correspondence and discussions. 
This research was supported in part by the Engineering and
Physical Sciences Research Council under Grant GR/J78327.

\end{multicols}


\begin{references}
%
\bibitem{Nien} B.~Nienhuis, in `Phase Transitions and Critical
Phenomena', vol.~11, C.~Domb and J.~L.~Lebowitz eds., Academic (1987).
%
\bibitem{DotFat} Vl.~S.~Dotsenko and V.~A.~Fateev, Nucl. Phys. B {\bf
240}, 312 (1984).
%
\bibitem{diFran} P.~di Francesco, H.~Saleur and J.-B.~Zuber,
J. Stat. Phys. {\bf 49}, 57 (1987).
%
\bibitem{JCcrossing} J.~L.~Cardy, J. Phys. A {\bf 25}, L201 (1992).
%
\bibitem{Wat} G.~Watts, J. Phys. A {\bf 29}, L363 (1996). 
%
\bibitem{logs} J.-S.~Caux, I.~Kogan and A.~M.~Tsvelik, Nucl. Phys. B
{\bf 466}, 444 (1996); 
V.~Gurarie, Nucl. Phys. B {\bf 546}, 774 (1999); 
J.~L.~Cardy, cond-mat/9911024.
%
\bibitem{GurLud} V.~Gurarie and A.~W.~W.~Ludwig, cond-mat/9911392.
%
\bibitem{Saltwist} H.~Saleur, J. Stat. Phys. {\bf 50}, 475 (1988).
%
\bibitem{JCtwist} J.~L.~Cardy, Nucl. Phys. B {\bf 275}, 200 (1986).
%
\bibitem{Gruz} I.~A.~Gruzberg, A.~W.~W.~Ludwig and N.~Read,
Phys. Rev. Let. {\bf 82}, 4254 (1999).
%
\bibitem{DSwinding} B.~Duplantier and H.~Saleur, Phys. Rev. Lett.
{\bf 60}, 2343 (1988).
%
\bibitem{sphere} The global topology should be thought of as that of a
sphere, so that a loop which winds around one point also winds around
the other.
%
\bibitem{JCarea} J.~L.~Cardy, Phys. Rev. Lett. {\bf 72}, 1580 (1994).

%
\bibitem{JCinprep} J.~L.~Cardy, in preparation.
%
\bibitem{nonleading} The leading twist operator for $Q=3$ is the (2,2)
operator. It becomes non-leading for $Q<2$. It counts hulls with a
weight $-\sqrt{Q'}$ rather than $+\sqrt{Q'}$.
%
\bibitem{Zirn} A.~Altland and M.~R.~Zirnbauer, Phys. Rev. B {\bf 55},
1142 (1997);
M.~R.~Zirnbauer, J. Math. Phys. {\bf 37}, 4986 (1996);
T.~Senthil, M.~P.~A.~Fisher, L.~Balents and C.~Nayak,
Phys. Rev. Lett. {\bf 81}, 4704 (1998);
T.~Senthil and M.~P.~A.~Fisher, Phys. Rev. B {\bf 60}, 6893 (1999);
V.~Kagalovsky, B.~Horovitz, Y.~Avishai and J.~T.~Chalker,
Phys. Rev. Lett. {\bf 82}, 3516 (1999); 
T.~Senthil, J.~B.~Marston and M.~P.~A.~Fisher, Phys. Rev. B {\bf 60},
4245 (1999);
J.~E.~Hirsch, Phys. Rev. Lett. {\bf 80}, 1834 (1999).
%
\bibitem{JCbound} J.~L.~Cardy, Nucl. Phys. B {\bf 324}, 581 (1989).
%
\bibitem{AAKK} This may be related to the observation in
Ref.~7 that correlators of the (1,3) operator with the Potts
energy operator (2,1) appear to be inconsistent when computed using the
degeneracy equations.
%
\end{references}
\end{document}